\documentclass{article}

\begin{document}

\begin{titlepage}

\title{Classical and Quantum Singularities of Levi-Civita Spacetimes with and without a Positive Cosmological Constant}       
\author{ D. A. Konkowski\thanks{email address: dak@usna.edu} \\Department of Mathematics \\ U.S. Naval Academy \\ Annapolis, Maryland, 21402 USA \and Cassidi Reese \\ Department of Physics \\ U.S. Naval Academy \\ Annapolis, Maryland, 21402 USA \and T.M. Helliwell\thanks{email address: T\_Helliwell@HMC.edu} and C. Wieland \\ Department of Physics \\ Harvey Mudd College \\ Claremont, California, 91711 USA}        
      
\maketitle
\end{titlepage}
\begin{abstract}
Levi-Civita spacetimes have classical naked singularities. They also have quantum singularities. Quantum singularities in general relativistic spacetimes are determined by the behavior of quantum test particles. A static spacetime is said to be quantum mechanically singular if the spatial portion of the wave operator is not essentially self-adjoint on a $C_{0}^{\infty}$ domain in $L^{2}$, a Hilbert space of square integrable functions. Here we summarize how Weyl's limit point-limit circle criterion can be used to determine whether a wave operator is essentially self-adjoint and how this test can then be applied to scalar wave packets in Levi-Civita spacetimes with and without a cosmological constant to help elucidate the physical properties of these spacetimes.  
\end{abstract}

\section{Introduction} 

Athough spacetime singularities have been studied since general relativity was first introduced, they are still not well understood \cite{TCE, berger, HE, ES}. There is still debate as to whether such singularities will exist in a unified theory (see, e.g., \cite{horowitz}). To begin understanding the effects of singularities in quantum gravity, it is important to probe classical singularities with quantum test wave packets and classical test fields \cite{wald, HM}. This is such an analysis for certain cylindrically symmetric spacetimes.

\par This conference proceeding starts with a review of the different properties of classical and quantum singularities. Next is a review of some mathematics, the Weyl limit point-limit circle criterion, that simplifies the evaluation of quantum singularities. The classical and quantum singularity structure of ordinary Levi-Civita spacetimes (without a cosmological constant) is then summarized. This is followed by a discussion of the classical and quantum singularity structure of Levi-Civita spacetimes with a positive cosmological constant. The paper ends with conclusions and a discussion of an area of further interest.

This conference proceeding is based in part on a paper by D.A. Konkowski, T.M. Helliwell and C. Wieland \cite{KHW} and in part on a senior thesis by C. Reese \cite{reese}.

\section{Singularities in General Relativity}

\subsection{Classical Singularities}   

In general relativity, a maximal spacetime is considered to be classically singular if the world line of a classical test particle ends after some finite proper time so that further evolution of the particle is not well defined \cite{HE}. This is usually summarized by saying that the maximal spacetime has incomplete geodesics and/or incomplete curves of bounded acceleration. 

\par A classification scheme devised by Ellis and Schmidt \cite{ES} divides classical singularities into three types: quasiregular, nonscalar curvature and scalar curvature. Quasiregular singularities are the mildest classical singularities and are associated with a topological obstruction such as the apex of a cone \cite{ES}. Even though an observer's world line would end at the singularity in some finite proper time, the observer would not see physical quantities diverge. Nonscalar curvature singularities are associated with finite curvature scalars but infinite tidal forces for particles that encounter it. This is due to the fact that curves exist through each point arbitrarily close to the nonscalar curvature singularity such that observers moving on these curves experience perfectly regular tidal forces. Scalar curvature singularities are the strongest of the classical singularities. They are associated with infinite curvature scalars such as at the center of a black hole or the beginning of a Big Bang cosmology. Physical quantities such as energy density and tidal forces diverge in the frames of all observers who approach these singularities. 

\par The three types of classical singularities can be expressed mathematically. A singular point is defined as the endpoint of an incomplete geodesics or incomplete curve of bounded acceleration in a maximal spacetime. A singular point $q$ is a quasiregular singularity if all components of the Riemann tensor are bounded at that point. If some components are not bounded, there exists a curvature singularity. A nonscalar curvature singularity exists if all scalars constructed from the metric, the totally antisymmetric tensor and the Riemann tensor tend to a finite limit. However, if any of these scalars are unbounded $q$ is a scalar curvature singularity.

\subsection{Quantum Singularitites}

To decide whether a spacetime is quantum mechanically singular we will 
use the
criterion proposed by Horowitz and Marolf \cite{HM} following early work by Wald \cite{wald} and Kay and Studer 
\cite{KS}. They call a spacetime 
quantum mechanically {\it non}singular if the evolution of a test wave 
packet 
in the spacetime is uniquely determined by the initial wave packet, without 
having to put arbitrary boundary conditions at the classical singularity. 
Their construction is restricted to static spacetimes.

According to Horowitz and Marolf, a static spacetime is quantum 
mechanically 
singular if the spatial portion of the Klein-Gordon wave operator is not 
essentially self-adjoint \cite{RS}. An operator, $L$, is called self-adjoint if

\begin{enumerate}
	\item[(i)] $L = L^{\dagger}$
	\item[(ii)] $Dom(L) = Dom(L^{\dagger})$
\end{enumerate}

\noindent where $L^{\dagger}$ is the adjoint of $L$. An operator is essentially self-adjoint if (i) is met and (ii) can be met by expanding the domain of the operator or its adjoint so that it is true \cite{RS}.

\par A relativistic scalar quantum particle 
with mass $M$ can be described by the positive frequency solution to the 
Klein-Gordon equation 

\begin{equation}
\frac{\partial^2\Psi}{\partial{t^2}}=-A\Psi
\end{equation}

\noindent in a static spacetime where the spatial operator 

\begin{equation}
A=-VD^{i}(VD_{i})+V^{2}M^{2} \nonumber
\end{equation}

\noindent with $V=-\xi_{\nu}\xi^{\nu}$. Here 
$\xi^{\nu}$ is the timelike Killing field and $D_{i}$ is the spatial 
covariant derivative on the static slice $\Sigma$. The Hilbert 
space is $L^{2}(\Sigma)$, the space of square integrable functions on 
$\Sigma$. 

If we initially define the domain of $A$ to be 
$C_{0}^{\infty}(\Sigma)$, $A$ is real, positive, symmetric operator and 
self-adjoint extensions always exist
\cite{RS}. If there is only a single, unique extension $A_{E}$, then $A$ is 
essentially self-adjoint. In this case, the Klein-Gordon equation for a free 
scalar particle takes the form \cite{HM}:  

\begin{equation}
i\frac{d\Psi}{dt}=A_E^{1/2}\Psi
\end{equation}

\noindent with 

\begin{equation}
\Psi(t)=exp(-it(A_E)^{1/2})\Psi(0). \nonumber\\
\end{equation}

These equations are ambiguous if $A$ is not essentially self adjoint. This 
fact led Horowitz and Marolf to define 
quantum mechanically singular spacetimes as those in which $A$ is not 
essentially self-adjoint. Examples are considered by Horowitz and Marolf
\cite{HM}, Kay and Studer \cite{KS}, Helliwell and Konkowski \cite{HK}, and Helliwell, Konkowski and Arndt \cite{HKA}.

\section{Mathematical Background}

\par  A particularly convenient way to establish essential
self-adjointness in the spatial operator of the Klein-Gordon equation
is to use the concepts of limit circle and limit point behavior.\footnote
{This section is based on {\bf Appendix to X.1} in Reed and
Simon \cite{RS}}  The
approach is as follows.  The Klein-Gordon equation for the 
spacetimes considered in this paper can be separated in the coordinates $t, r, \theta, z$.
Only the radial equation is non-trivial.  With changes in both
dependent and independent variables, the radial equation can be
written as a one-dimensional Schr\"{o}dinger equation

\begin{equation}
    H\Psi(x) = E\Psi(x)
    \label{eq:13}
\end{equation}

\noindent where $x \in (0,\infty )$ and the operator  $H = - d^{2}/dx^{2}
+ V(x)$.

\newtheorem{Definition}{Definition}
\begin{Definition} The potential $V(x)$ is in the limit circle case
at $x = 0$  if for some, and therefore for all $E$, {\it all}
solutions of equation (\ref{eq:13}) are square integrable at zero.  If
$V(x)$  is not
in the limit circle case, it is in the limit point case.
\end{Definition}

\par  A similar definition pertains to $x=\infty$.  The potential
$V(x)$ is in the limit circle case at $x=\infty$ if all solutions of
equation (\ref{eq:13}) are square integrable at infinity; otherwise,
$V(x)$ is in
the limit point case at infinity.

\par There are of course two linearly independent solutions of the
Schr\"{o}dinger equation for given $E$.  If $V(x)$ is in the limit circle
case at zero, both solutions are $\mathcal{L}^{2}$ at zero, so all linear
combinations are $\mathcal{L}^{2}$ as well.  We would therefore need a
boundary
condition at $x=0$ to establish a unique solution.  If $V(x)$ is in
the limit {\it point} case, the $\mathcal{L}^{2}$ requirement eliminates
one of
the solutions, leaving a unique solution without the need of
establishing a boundary condition at $x=0$. This is the whole idea of
testing for quantum singularities; there is no singularity if the
solution is unique, as it is in the limit point case.  The critical
theorem is due to Weyl \cite{RS, weyl}.

\newtheorem{theorem}{Theorem}
\begin{theorem}[The Weyl limit point-limit circle criterion.] If
$V(x)$ is a continuous real-valued function on $(0, \infty)$, then
$H = - d^{2}/dx^{2} + V(x)$ is essentially self-adjoint on
$C_{0}^{\infty}(0, \infty)$ if and only if $V(x)$ is in the limit
point case at both zero and infinity.
\end{theorem}

\par The following theorem can be used to establish the limit circle-limit
point behavior at infinity \cite{RS}.

\newtheorem{theorem1}[theorem]{Theorem}
\begin{theorem1}[Theorem X.8 of Reed and Simon \cite{RS}.] If $V(x)$ is
continuous and real-valued on $(0, \infty)$, then $V(x)$ is in the
limit point case at infinity if there exists a {\em positive}
differentiable function $M(x)$ so that
\begin{enumerate}
    \item[(i)] $V(x) \ge - M(x)$
    \item[(ii)] $\int_{1}^{\infty} [M(x)]^{-1/2} dx = \infty$
    \item[(iii)] $M'(x)/M^{3/2}(x)$ is bounded near $\infty$.
\end{enumerate}
Then $V(x)$ is in the limit point case (complete) at $\infty$.
\end{theorem1}

A sufficient choice of the $M(x)$ function for our purposes is the
power law function $M(x) = c x^{2}$ where $c > 0$. Then {\it (ii)} and
{\it (iii)} are satisfied, so if $V(x) \ge -c x^{2}$, $V(x)$ is in the
limit point case at infinity.

\par A theorem useful near zero is the following.

\newtheorem{theorem2}[theorem]{Theorem}
\begin{theorem2} [Theorem X.10 of Reed and Simon \cite{RS}.]
Let $V(x)$ be continuous and {\it positive} near zero.
If $V(x)\ge\frac{3}{4} x^{-2}$ near zero then $V(x)$ is in the limit
point case.  If for some $\epsilon > 0$,
$V(x)\le(\frac{3}{4}-\epsilon)x^{-2}$ near zero, then $V(x)$ is in the
limit circle case.
\end{theorem2}

\noindent These results can now be used to help test for quantum
singularities in the Levi-Civita spacetimes.

\section{Ordinary Levi-Civita Spacetimes}

The metric for an ordinary Levi-Civita (LC) spacetime \cite{LC} (no cosmological constant) has the form

\begin{equation}
    ds^{2} = -r^{4 \sigma}dt^{2} + r^{8 \sigma^{2} + 4\sigma}(dr^{2}+
    dz^{2}) + \frac{r^{2 - 4 \sigma}}{C^{2}} d\theta^{2}
    \label{eq:6}
\end{equation}

\noindent where $\sigma$ and $C$ are real numbers ($C>0$) and the coordinates are cylindrical with the usual ranges. For some
parameter values one can interpret the Levi-Civita spacetime as the
spacetime of an ``infinite line mass". In fact, after some controversy
in the literature (see, e.g. \cite{bonnor}, \cite{HRS}, \cite{HSTW}), the
following interpretations have become somewhat accepted: $\sigma = 0, 1/2$
locally flat; $\sigma =0,\,\ C=1$ Minkowski spacetime; $\sigma =0,\,\ C\not= 1$
cosmic string spacetime; $0 < \sigma < 1/2$ ``infinite line mass" spacetime
(modeled by a scalar curvature singularity at $r=0$); $\sigma = 1/2$ Minkowski 
spacetime in accelerated coordinates (planar source). 

\par The following discussions of the classical and quantum singularities in LC spacetimes is based on the paper by Konkowski, Helliwell and Wieland \cite{KHW}.

\subsection{Classical Singularities}

Computation of the Kretschmann scalar

\begin{equation}
    R_{\mu \nu \sigma \tau}R^{\mu \nu \sigma \tau} = \frac{64
    \sigma^{2} (2\sigma -1)^{2}}{(4 \sigma^{2} -2\sigma +1)^{3}\,\ r^{4}}
    \label{eq:7}
\end{equation}

\noindent shows that $r=0$ is a scalar curvature singularity for all
$\sigma $
except $\sigma = 0, 1/2$. It can be shown that all 14 scalar polynomial invariants in the curvature have the same properties. In fact, the spacetimes with $\sigma= 0, 1/2$ are each flat.

\par If $\sigma = 0$, the metric  
\begin{equation}
    ds^{2} = -dt^{2} + dr^{2} +dz^{2} + \frac{r^{2}}{C^{2}} d \theta^{2}.
    \label{one}
\end{equation}

\noindent If $C=1$, this is simply Minkowski spacetime in
cylindrical coordinates.  If $C \not= 1$, equation (\ref{one}) is the
metric for an idealized cosmic string. There is a quasiregular
(``disclination'') singularity at $r=0$ (see, e.g., \cite{HKA} for a
discussion). This is a
topological singularity, not a curvature singularity, and the parameter C describes a topological property of the spacetime, its deficit angle.

\par The $\sigma=1/2$ metric

\begin{equation}
    ds^{2} = -r^{2} dt^{2} + dr^{2} + dz^{2} + \frac{1}{C^{2}}
    d\theta^{2}
    \label{eq:10}
\end{equation}

\noindent is also flat but its interpretation is more difficult \cite{bonnor, HRS, HSTW}. This metric can be transformed to Minkowski coordinates

\begin{equation}
    ds^{2} = -d{\bar t}^{2} + d{\bar x}^{2} + d{\bar y}^{2} + d{\bar
    z}^{2}
    \label{eq:11}
\end{equation}

\noindent where ${\bar t} = r \sinh t$, ${\bar x} = r \cosh t$, ${\bar y} = \theta/C$,
and ${\bar z} = z$ and the ${\bar y}$ coordinate can now range
from $-\infty$ to $\infty$. This is flat spacetime described from the point of
view of an accelerating frame of reference. This seems to support an
interpretation of the $\sigma = 1/2$ case as a planar source producing
flat spacetime described by a uniformly accelerating observer
\cite{bonnor, GH}. In other words, one can interpret it as the spacetime of a
gravitational field produced by an
infinite planar sheet of positive mass density.

\subsection{Quantum Singularities}
The analysis in \cite{KHW} uses Weyl's
limit point-limit circle criterion to determine essential
self-adjointness of the spatial portion of the Klein-Gordon
wave operator on a $C_{0}^{\infty}(\Sigma)$ domain in $L^{2}(\Sigma)$, a Hilbert space of square integrable functions.
The conclusions will now be summarized.

If $\sigma$ is neither zero nor one-half, the
Klein-Gordon operator is not
essentially self-adjoint,  so all $\sigma \neq
0,\,\ \sigma \neq 1/2$ Levi-Civita spacetimes are quantum mechanically
singular as well as being classically singular with scalar curvature singularities.

\par If $\sigma = 0$ and $C = 1$, the spacetime is simply Minkowski
space. One of the two solutions of the radial Klein-Gordon equation
can be rejected because it diverges at a regular point ($r=0$) of the
spacetime. The operator is therefore quantum
mechanically nonsingular (a well known fact, repeated here for
completeness).

\par If $\sigma = 0$ and $C \neq 1$, the spacetime is the conical
spacetime corresponding to an idealized cosmic string. The cosmic string spacetimes
are quantum mechanically
singular for azimuthal quantum number $m$ such that $|m| C < 1$ and
nonsingular if  $|m| C \ge 1$. If arbitrary values of $m$ are allowed,
these
spacetimes are quantum mechanically singular in agreement with earlier
results \cite{HK}. These spacetimes are also
classically singular with a quasiregular (``disclination'')
singularity at $r=0$.

\par If $\sigma = 1/2$ the classical spacetime is flat and without a
classical singularity. This spacetime is also quantum mechanically nonsingular.
The Weyl limit point-limit circle
techniques used in \cite{KHW} emphasize the flatness of the spacetime
and support a description given in \cite{bonnor} of this spacetime as one
given by a cylinder whose  radius has tended to infinity.

\par  For the Levi-Civita spacetimes, all that are classically
singular are also generically quantum mechanically singular, and all that are
classically nonsingular ($\sigma = 0,\,\ C = 1$, and $\sigma = 1/2$) are
also quantum mechanically nonsingular.  The classically and
quantum-mechanically nonsingular spacetimes correspond to isolated
values of $\sigma$, so that (for example) even though the spacetime
$\sigma = 0,\,\ C = 1$ is nonsingular, the spacetimes with  $\sigma
\to 0,\,\ C = 1$ are singular.  The only discrepency between
classical and quantum singularities are for the $\sigma =0,\,\ C \neq 1$
modes with $|m| C \ge 1$, which produce no quantum singularity
in a classically singular spacetime. The physical reason is that the
wavefunction for large values of $m$ in a flat space with a quasiregular
singularity at $r=0$ is unable to detect the presence of the
singularity because of a repulsive centrifugal potential.

\section{Levi-Civita Spacetimes with Positive Cosmological Constant}

Levi-Civita spacetimes with a cosmological constant (LCC) have been studied by de Silva et al \cite{deSilva}. Originally placed into the field equations as a constant to ensure a static universe, the possibility of a positive cosmological constant is gaining much popularity as it could account for the acceleration of the expansion of the universe as supported by recent astronomical observations of type Ia supernovae \cite{riese}.  In general, the metric for LCC spacetimes can be written as 

\begin{eqnarray}
ds^{2}&=&-Q(r)^{2/3}(P(r)^{-2(4\sigma^{2}-8\sigma+1)/3\Lambda}dt^{2}\nonumber\\ 
&+&P(r)^{2(8\sigma^{2}-4\sigma-1)/3\Lambda}dz^{2}+ C^{-1}P(r)^{-4(2\sigma^{2}+2\sigma-1)/3\Lambda}d\phi^{2})+ dr^{2}
\end{eqnarray}

\noindent where $\{x^{\mu}\} = \{t,r,\theta,\phi\}$ are the usual cylindrical coordinates with the usual ranges and $A=4\sigma^{2}-2\sigma+1$. The constant  $C$ is related to angle defects and the constant $\sigma$ is related to mass per unit length. The functions $P(r)$ and $Q(r)$ are defined as

\begin{equation}
P(r)=\frac{2}{\sqrt{3\Lambda}}\tan(\frac{\sqrt{3\Lambda r}}{2}),\,\
Q(r)=\frac{1}{\sqrt{3\Lambda}}\sin(\sqrt{3\Lambda r})
\end{equation}

\noindent where $\Lambda$ is the cosmological constant. Here we will only consider the cases where $\Lambda>0$ to correspond to the apparent physical reality of an expanding and accelerating universe. From equations (11) and (12), it can be shown that as $r\rightarrow 0,\,\ Q(r)\approx r,\,\ P(r)\approx r$ and ordinary Levi-Civita spacetime is regained, the results of whose study \cite{KHW} were summarized in Section 4 and will not be discussed further here.

\par However, equation (11) is also singular on the hypersurface $r=r_{g}=\pi/\alpha$, where $\alpha\equiv(3|\Lambda|)^{1/2}$. As $r\rightarrow r_{g},\,\ Q(r) \approx R$ and $P(r) \approx R^{-1}$, where $R\equiv r-r_{g}$. The approximate metric is

\begin{equation}
ds^{2}\approx -R^{4(4\sigma^{2}-5\sigma+1)/3A}dt^{2}+R^{-4(2\sigma^{2}-\sigma-1)/3A}dz^{2}+C^{-2}R^{2(8\sigma^{2}+2\sigma-1)/3A}d\phi^{2}+dR^{2} \label{three}
\end{equation}

\noindent where $R\approx 0$. For the sake of mathematical convenience, the approximation in equation (13) will be considered exact throughout the rest of this paper. Although this is not valid away from the hypersurface, the behavior of interest occurs near $R\approx0$, where the equality holds true. In this case, the spacetime described by equation (\ref{three}) generically contains a classical timelike scalar curvature singularity at $R=0$. In some cases we will consider $R=0$ to be the symmetry axis rather than $r=0$ \cite{deSilva}. Those cases will be specially noted.

\par The following discussion of the classical and quantum singularities in LCC spacetimes is based on the senior thesis of C. Reese \cite{reese}.

\subsection{Classical Singularities}

Computation of the Kretschmann scalar

\begin{equation}
    R_{\mu \nu \sigma \tau}R^{\mu \nu \sigma \tau} = \frac{64
    (\sigma -1)^{2}(2\sigma+1)^{2}(4\sigma-1)^{2}}{27 A^{3} R^{4}}
    \label{eq:7}
\end{equation}

\noindent shows that $R=0$ is a scalar curvature singularity for all
$\sigma $
except $\sigma = -1/2, 1/4,1$. It can be shown that all 14 scalar polynomial invariants in the curvature have the same properties. In fact, the spacetimes with $\sigma = -1/2, 1/4, 1$ are each flat.

\par If $\sigma = -1/2$, the metric  
\begin{equation}
    ds^{2} = -R^{2}dt^{2} + dR^{2} +dz^{2} + C^{-2} d \phi^{2}.
    \label{eq:8}
\end{equation}

\noindent This solution is flat and has planar symmetry. It is similar to the $\sigma=1/2$ case for the ordinary LC spacetime; that is, it is flat spacetime described from the point of view of an accelerating reference frame.

\par If $\sigma = 1/4$, the metric  
\begin{equation}
    ds^{2} = -dt^{2} + dR^{2} +R^{2}dz^{2} + C^{-2} d \phi^{2}.
    \label{eq:8}
\end{equation}

\noindent This solution is flat. It resembles Minkowski spacetime in cylindrical coordinates with the variable $z$ acting as the angular variable. Applying the coordinate transformation $\bar{\phi}= C^{-1}\phi$, where $-\infty<\bar{\phi}<\infty$, and  $x=R\cos(z)$ and $y=R\sin(z)$, equation (16) becomes

\begin{equation}
    ds^{2} = -dt^{2} + dx^{2} +dy^{2} +  d \bar{\phi}^{2}.
    \label{eq:8}
\end{equation}

\noindent which is ordinary Minkowski spacetime. Thus $R=0$ is simply a coordinate singularity in this case.

\par If $\sigma = 1$, the metric  
\begin{equation}
    ds^{2} = -dt^{2} + dR^{2} +dz^{2} + C^{-2} d \phi^{2}.
    \label{eq:8}
\end{equation}

\noindent If $C=1$, this is simply Minkowski spacetime in
cylindrical coordinates.  If $C \not= 1$, equation (18) is the
metric for an idealized cosmic string (assuming we let $R=0$ be the symmetry axis of the spacetime). There is a quasiregular
(``disclination'') singularity at $R=0$ (see, e.g., \cite{HKA} for a
discussion). This is a
topological singularity, not a curvature singularity, and the parameter C describes a topological property of the spacetime, its deficit angle. This case is similar to the $\sigma=0$ case for the ordinary LC spacetime.

\subsection{Quantum Singularities}

The existence of quantum singularities is next considered. Applying the Klein-Gordon equation to the LCC metric and assuming a solution of the form, \\
$\Phi\sim e^{-i\omega t}e^{im\phi}e^{ikz}h(R)$, the following radial equation is obtained:

\begin{eqnarray}
h''(R)+\frac{1}{R}h'(R)&+& [\omega^{2}R^{-4(4\sigma^{2}-5\sigma+1)/3A}\nonumber\\
&-& m^{2} C^{2}R^{-2(8\sigma^{2}+2\sigma-1)/3A}\nonumber\\
&-&k^2 R^{4(2\sigma^{2}-\sigma-1)/3A}-M^{2}]h(R)=0.
\end{eqnarray}

\noindent In order to apply the limit point-limit circle criterion from Section 3, the radial equation must first be changed to a one-dimensional Schr\"{o}dinger-equation form

\begin{equation}
\frac{d^{2}\Psi}{dx^{2}} + [E - V(x)] \Psi = 0
\end{equation}

\noindent where $E$ is a kinetic energy term that is constant for the system and $V(x)$ is a potential energy term dependent upon the position. Except for the special case $\sigma = -1/2$, the substitutions $h(R) = x^{-1/2}\Psi(x)$ and $R = (x^2 \alpha C)^{1/2\alpha}$, where $\alpha\equiv(2\sigma + 1)^{2}/3A$ may be used to transform equation (19) to equation (20). If $E=\omega^{2}\alpha^{-1}C$ and

\begin{eqnarray}
V(x)&=&\frac{m^{2}C^{3}}{\alpha} (\alpha C x^{2})^{(-4 \sigma +1)/\alpha A}\nonumber\\
&+& \frac{M^{2}C}{\alpha}(\alpha C x^{2})^{2(4\sigma^{2} - 5 \sigma + 1)/3\alpha A}\nonumber\\
&+& \frac{k^{2}C}{\alpha}(\alpha C x^{2})^{4\sigma(\sigma -1)/\alpha A}
\nonumber\\&-&\frac{1}{4}x^{-2},
\end{eqnarray}

\noindent the form of equation (20) is obtained.

\subsubsection{The Generic Case}

\par In order to determine the limit point-limit circle behavior near infinity, Theorem 2 of Section 3 must be applied. A choice of $M(x)=cx^{2}$ where $c>0$ satisfies {\it(ii)} and {\it(iii)}.  For the potential given by equation (21) as      $x \rightarrow \infty$, {\it(i)} will also be satisfied. Thus, $V(x)$ is always in the limit point case at infinity for all values of $\sigma$, except $\sigma=-1/2$, as the substitutions do not apply in this case.

\par To study the behavior in the limit $x \rightarrow 0$, the important terms in $V(x)$ must be identified. The first and third terms are never more divergent than $x^{-2}$, with the minimum values occurring at $\sigma =1$ and $\sigma =1/4$, respectively. The second term diverges as $x^{-1/2}$, with a minumum at $\sigma=-1/2$. The last term always diverges as $x^{-2}$. The three cases of $\sigma$ that make any of the first three terms significant are also those which are classically important as they are the values which lead to a non-divergent Kretschmann scalar. It is interesting to note that these values are also special when determining the quantum mechanical nature of the singularities in the spacetime. Thus, for all values of $\sigma$ except $\sigma= -1/2, 1/4, 1$, equation (20) has the form

\begin{equation}
\frac{d^{2}\Psi}{dx^{2}}+\frac{1}{4x^{2}}\Psi = 0
\end{equation}

\noindent as $x \rightarrow 0$. The solutions are $\Psi_{1} = x^{1/2}$ and $\Psi_{2} = x^{1/2}\ln(x)$, both of which are square integrable at $x=0$. Thus according to Theorem 3, the potential is in the limit circle case. 

\par Since $V(x)$ is not in the limit point case at both zero and infinity, it is not essentially self-adjoint and the spacetime is quantum mechanically singular. Thus, the LCC spacetimes with a classical scalar curvature singularity at $R=0$ are also quantum mechanically singular.

\subsubsection{Special Cases}

\par Three special cases were found for this metric wherein the curvature scalars do not diverge as $R\rightarrow0$. They are $\sigma= -1/2, 1/4, 1$. It was determined in Section 4 that the potential is always in the limit point case at infinity, for all $\sigma$ except $\sigma = -1/2$. Thus, only the behavior near zero needs to be studied for the other two special cases $\sigma=1/4, 1$; however, for $\sigma =-1/2$, the behavior near both zero and infinity needs to be studied once the new potential is obtained. 

\subsubsection*{\bf Case(1)\,\ $\sigma = -1/2$}

\par For this case, new substitutions for $h(R)$ and $R$ must be chosen to find the potential, equation (19) becomes

\begin{equation}
h''(R)+\frac{1}{R} h'(R) + [ \omega^{2}R^{-2} - m^{2} C^{2} - k^{2} - M^{2}] h(R) =0.
\end{equation}

\noindent Using the substitutions $h(R) = \Psi(x)$ and $R=e^{Cx}$, equation (23) transforms to equation (20) with $E=\omega^{2}C^{2}$ and

\begin{equation}
V(x) = m^{2}C^{4}e^{2Cx} + k^{2}C^{2}e^{2Cx} +M^{2}C^{2}e^{2Cx}.
\end{equation}

Now $R=0$ corresponds to $x\rightarrow -\infty$ and $R \rightarrow \infty$ corresponds to $ x \rightarrow \infty$. Thus the behavior at $x=\pm\infty$ needs to be studied, and Theorem 2 to Section 3 must be employed. A choice of $M(x) = - \beta x^{2}$, where $\beta >0$ will satisfy conditions {\it (ii)} and {\it (iii)}. Since all terms in equation (24) are positive, this choice will also satisfy condition {\it (i)}. Thus, for the case of $\sigma = -1/2$, the potential is in the limit point case at $x=\pm\infty$, corresponding to $R=0$ and $R=\infty$. The radial operator is thus essentially self-adjoint and we can call the spacetime quantum mechanically nonsingular. This is the planar case that is very similar to the $\sigma=1/2$ case in the ordinary LC spacetime.

\subsubsection*{\bf Case(2)\,\ $\sigma=1/4$}

In this case, as $x\rightarrow 0$, $V(x) \rightarrow (k^{2} - 1/4)x^{-2}$, and thus equation (20) has the form

\begin{equation}
\frac{d^{2}\Psi}{dx^{2}} - (\frac{k^2-1/4}{x^{2}})\Psi = 0.
\end{equation}

\noindent For $k \ne 0$, the solutions are $\Psi_{1} = x^{1/2 + |k|}$ and $\Psi_{2} = x^{1/2 - |k|}$, and for $k=0$, the solutions are $\Psi_{1} = x^{1/2}$ and $\Psi_{2} = x^{2}\ln{x}$. Thus, $\Psi_{1}$ is always square integrable, but $\Psi_{2}$ is only square integrable when $k=0$. However, since $R=0$ is a regular point of the spacetime, solutions which diverge at $R=0$ must be discarded and, therefore, there is only one square integrable $\Psi$ for each $k$ value. The radial operator is thus essentially self-adjoint as it must be in a complete Minkowski spacetime, and the $\sigma=1/4$ spacetime is quantum mechanically nonsingular.

\subsubsection*{\bf Case(3) \,\ $\sigma=1$}

In this case, as $x\rightarrow 0$, $V(x) \rightarrow (m^{2} C^{2} - 1/4)x^{-2}$, and thus equation (20) has the form

\begin{equation}
\frac{d^{2}\Psi}{dx^{2}} - (\frac{m^{2}C^2-1/4}{x^{2}})\Psi = 0.
\end{equation}

\noindent In a manner similar to the previous case, for $m \ne 0$, the solutions are $\Psi_{1} = x^{1/2 + |m|C}$ and $\Psi_{2} = x^{1/2 - |m|C}$, and for $m=0$, the solutions are $\Psi_{1} = x^{1/2}$ and $\Psi_{2} = x^{2}\ln{x}$. Thus, $\Psi_{1}$ is always square integrable, but $\Psi_{2}$ is only square integrable when $|m|C < 1$. However, unlike the previous case, $R=0$ is not a regular point of the spacetime in all cases; it is only a regular point if $C=1$, otherwise it is a quasiregular singular point. Therefore, only in the $C = 1$ case can we discard the solutions which diverge at $R=0$ and automatically obtain a quantum mechanically nonsingular spacetime; it is simply a complete Minkowski spacetime. The other cases are more complicated: When $C \ne 1$, $V(x)$ is in the limit point case if and only if $|m|C \ge 1$ and in the limit circle case if and only if $|m|C < 1$. There is a range of $m$ in each quasiregular case which cause a quantum singularity to occur as in the $\sigma = 0$ LC spacetime. If this range of modes is forbidden the classically singular spacetime is quantum mechanically nonsingular, but, in the generic case, LCC spacetimes with quasiregular singularities are quantum mechanically singular.

\section{Conclusions and an Area of Further Interest}

\par  For the LC and LCC spacetimes, all that are classically
singular are also generically quantum mechanically singular, and all that are
classically nonsingular are also quantum mechanically nonsingular. The only discrepency between
classical and quantum singularities are (1) for the $\sigma =0,\,\ C \neq 1$
modes with $|m| C \ge 1$ in the LC case, and (2) for the $\sigma =1,\,\ C \neq 1$
modes with $|m| C \ge 1$ in the LCC case; each of these sets of modes produces no quantum singularity in a classically singular spacetime. The physical reason is that the
wavefunction for large values of $m$ in a flat space with a quasiregular
singularity is unable to detect the presence of the
singularity because of a repulsive centrifugal potential.  

\par Finally, there is an area of further interest. The importance of the underlying Hilbert space should be considered through a comparison in this instance of the notion of quantum singularity with the notion of wave singularity \cite{IH}. In the latter, the Hilbert space is the first Sobolev space $H^{1}$ rather than the space of square integrable functions $L^{2}$. Spacetimes that are quantum mechanically singular may be wave regular \cite{IH}. We are currently studying LC and LCC spacetimes for wave regularity, and plan to address it, together with a related notion of global hyperbolicity \cite{vickers}, in an upcoming paper \cite{KHR}.

\section{Acknowledgements}

One of us (DAK) was partially funded by NSF grants PHY-9988607 and PHY-02411384 to the U.S. Naval Academy. She also thanks Queen Mary, University of London, where some of this work was carried out.


\begin{thebibliography}{0}

\bibitem{TCE} F. Tipler, C. Clarke, and G. Ellis, "Singularities and Horizons: A Review Article," in A. Held {\it General Relativity and Gravitation, Volume 2} (New York: Plenum Press, 1980) 97.

\bibitem{berger} B.K. Berger, "Numerical Approaches to Spacetime Singularities," {\it Living Reviews in Relativity}, No. 2002-1.

\bibitem{HE} S.W. Hawking and G.F.R. Ellis, {\it The Large-Scale
     Structure of Spacetime} (Cambridge: Cambridge University Press, 1973).

\bibitem{ES} G.F.R. Ellis and B.G. Schmidt, {\it Gen. Relativ. Grav}
    {\bf 8}, 915 (1977).

\bibitem{horowitz}G.T. Horowitz and J. Polchinski, {\it Phys. Review D} {\bf 66}, 103512 (2002).

	
\bibitem{wald} R.M. Wald, {\it J. Math Phys.} {\bf 21}, 2802 (1980).


\bibitem{HM} G.T. Horowitz and D. Marolf, {\it Phys. Rev. D} {\bf 52}, 5670 (1995).

\bibitem{KHW} D.A. Konkowski, T.M. Helliwell and C. Wieland, {\it Class. Quantum Grav.} {\bf 21}, 265 (2004).

\bibitem{reese} C. Reese, "Classical and Quantum Singularities in Levi-Civita Spacetimes with Cosmological Constant," {\it Proceedings of the National Conference on Undergraduate Research (NCUR)} 2004.

\bibitem{RS} M. Reed and B. Simon {\it Functional Analysis}
    (New York: Academic Press,1972); M. Reed and B. Simon 1972 {\it Fourier Analysis and
    Self-Adjointness} (New York: Academic Press, 1972)


\bibitem{HKA} T.M. Helliwell, D.A. Konkowski and V. Arndt, {\it Gen. Relativ. Grav.} {\bf 35}, 79 (2003).

\bibitem{HK} D.A. Konkowski and T.M. Helliwell, {\it Gen.
     Relativ. Grav.} {\bf 33}, 1131 (2001).

\bibitem{weyl} H. Weyl {\it Math. Ann.} {\bf 68} 220 (1910).

\bibitem{LC} T. Levi-Civita {\it Rend. Acc. Lincei} {\bf 28}
    101 (1919).

\bibitem{bonnor} W.B. Bonnor ``The Static Cylinder in General Relativity''
    in {\it On Einstein's Path}  ed. A. Harvey (New York:
    Springer, 1999) 113.

\bibitem{HRS} L. Herrera, J. Ruifern\'{a}ndez and N.O. Santos {\it Gen.
    Relativ. Grav.} {\bf 33} 515 (2001).

\bibitem{HSTW} L. Herrera, N.O. Santos, A.F.F. Teixeira, and A.Z. Wang {\it Class. Quantum Grav.} {\bf 18} 3847 (2001).

\bibitem{GH} R. Gautreau and R.B. Hoffmann, {\it Nuovo Cimento B}
    {\bf 61} 411 (1969).


\bibitem{deSilva} M.F.A. de Silva, A. Wang, F.M. Paiva and N.O. Santos, {\it Phys. Review D} {\bf 61}, 44003 (2000).

\bibitem{riese} A.G. Riese {\it et al}, {\it Astron. J.} {\bf 116}, 1009 (1998).

    
\bibitem{KS} B.S. Kay and U.M. Studer, {\it Commun. Math. Phys.}
    {\bf 139}, 103 (1991).

\bibitem{IH} A. Ishibashi and A. Hosoya, {\it Phys. Rev. D} {\bf 60}, 104028 (1999).

\bibitem{vickers} J.A. Vickers and J.P. Wilson, {\it Class. Quantum Grav.} {\bf 17} 1333 (2000).

\bibitem{KHR} T.M. Helliwell, D.A. Konkowski and C. Reese, in preparation.


\end{thebibliography}
\end{document}